\documentclass[twocolumn,showpacs,preprintnumbers,amsmath,amssymb,ctexart]{revtex4}
\usepackage{pifont}
\usepackage{CJK}
\usepackage{tipa}
\usepackage{graphicx}
\usepackage{dcolumn}
\usepackage{slashbox}
\usepackage{bm}
\usepackage{multirow}

\begin{document}
\begin{CJK*}{GBK}{}
\CJKtilde
\title{Study of radially excited $D_s(2^1S_0)$ and $D^+_{sJ}(2632)$ }

\author{Yu Tian, Ze Zhao
and Ailin Zhang\footnote{Corresponding author:
zhangal@staff.shu.edu.cn}} \affiliation{Department of Physics,
Shanghai University, Shanghai 200444, China}
\begin{abstract}
The $J^P=0^-$ radial excitation $D_s(2^1S_0)$ is anticipated to have mass $2650$ MeV (denoted with $D_s(2650)$) though it has not been observed. $D_s(2650)$ is anticipated to be observed in inclusive $e^+e^-$ and $pp$ collisions in $D^*K$ channel. $D_s(2650)$ can be produced from hadronic decays of higher excited resonances. In a $^3P_0$ model, hadronic production of $D_s(2650)$ has been studied, relevant decay widths have been estimated. The hadronic decays of $D_s(3P)$ have been explored in detail, and the dominant decay channels have been pointed out. Two hadronic decay channels of $D^+_{sJ}(2632)$ were observed by SELEX, but have not been observed by any other experiment. Hadronic decays of $D^+_{sJ}(2632)$ in different assignments have been explored in the $^3P_0$ model. In view of the hadronic decay manner of $D^+_{sJ}(2632)$ , we conclude that $D^+_{sJ}(2632)$ cannot be a conventional charmed strange $c\bar s$ meson.
\end{abstract}

\maketitle
\end{CJK*}

\section{Introduction}
The lowest lying $1S$ and $1P$ excited $D$ and $D_s$ states are believed established~\cite{pdg,gi}. In recent years, more and more higher excited $D$ and $D_s$ resonances have been observed though some of them have not been identified. Study of their spectroscopy, production and decay will be helpful for their identification and quark dynamics.

For $D$ resonances, $D(2550)$, $D^*(2600)$, $D(2750)$ and $D^*(2760)$ were observed in inclusive $e^+e^-$ and $pp$ collisions by BaBar~\cite{amo} and LHCb~\cite{lhcb1} collaborations, respectively. $D^*(2760)$ was subsequently found consisting of $D^*_1(2760)$ and $D^*_3(2760)$~\cite{lhcb2,lhcb3}.

For $D_s$ resonances, $D^*_{s1}(2700)$ and $D^*_{sJ}(2860)$ were observed by Belle~\cite{belle} and BaBar~\cite{babar} collaborations. $D^*_{sJ}(2860)$ was also found consisting of $D^*_{s1}(2860)$ and $D^*_{s3}(2860)$~\cite{lhcb4,lhcb5}. According to Refs.~\cite{colangelo,gm,liu3,zhong,wang,chen2,chen,zhang3,zhang4,zhu}, we have a tentative assignment of these $D$ and $D_s$ resonances given in Table~\ref{tab1}

\begin{table}[t]\label{tab1}
\caption{Tentative assignment of $2S$ and $1D$ excited $D$ and $D_s$ resonances.}
\begin{tabular}{p{0.0cm}p{1.4cm}p{1.4cm}p{1.4cm}p{1.4cm}p{1.4cm}p{1.4cm}}
   \hline\hline
   & $\setminus n^{2S+1}L_J$ & $2^1S_0$     & $2^3S_1$           & $1^{1,3}D_2$  & $1^3D_1$          & $1^3D_3$\\
   \hline
   &$D$                      & $D(2550)$    & $D^*(2600)$        & $D(2750)$     & $D^*_1(2760)$     & $D^*_3(2760)$\\
   &$D_s$                    & $--$  & $D^*_{s1}(2700)$          & $--$          & $D^*_{s1}(2860)$  & $D^*_{s3}(2860)$\\
   \hline\hline
\end{tabular}
\label{masses}
\end{table}

When these highly excited $D$ and $D_s$ resonances are examined, it is interesting to note that the masses of $D^*_{s1}(2700)$, $D^*_{s1}(2860)$ and $D^*_{s3}(2860)$ are about $100$ MeV higher than those of $D^*(2600)$, $D^*_1(2760)$ and $D^*_3(2760)$, respectively. Therefore, two missing $D_s$ with masses $2650$ MeV and $2850$ MeV are expected to correspond to the observed $D(2550)$ and $D(2750)$. In Ref.~\cite{chen2}, $D_{sJ}(2850)^\pm$ has been predicted with $\Gamma_{total}=125.1$ MeV and $\Gamma_{total}=51.6$ MeV corresponding to $(J^P=2^-,j_q={3\over 2})$ and $(J^P=2^-,j_q={5\over 2})$, respectively.

$D(2550)$ was suggested as the $J^P=0^-$ radial excitation $D(2^1S_0)$~\cite{wang,chen2,colangelo,gm}, and $D(2750)$ was suggested as the $J^P=2^-$ admixture of $D(1^1D_2)$ and $D(1^3D_2)$~\cite{chen2,gm,zhong}. Therefore, the two missing $D_s$ are expected to have $J^P=0^-$ and $J^P=2^-$. In this paper, the $D_s(2^1S_0)$ with mass around $2650$ MeV will be denoted as $D_s(2650)$ and be studied. The $J^P=2^-$ $D_s$ resonance will not be addressed for an unclear mixing.

In $2600\sim 2700$ MeV energy region, a surprisingly narrow charmed strange state, $D^+_{sJ}(2632)$, was first reported by SELEX~\cite{SELEX} in $D^+_s\eta$ and $D^0K^+$ decay channels: $D^+_{sJ}(2632)\rightarrow D^0K^+, D^+_s\eta$. $D_{sJ}(2632)$ was observed with mass $M=2632.5\pm1.7$ MeV and $\Gamma<17$ MeV. In particular, it has an exotic relative branching ratio $\Gamma(D^0K^+)/\Gamma(D^+_s\eta)=0.14\pm0.06$. However, subsequent search for $D^+_{sJ}(2632)$ by BaBar~\cite{babar2}, FOCUS or BELLE has not found any evidence.

Since the report of $D^+_{sJ}(2632)$, there appear many interpretations of this state. In Refs.~\cite{chao,ted,rupp}, $D^+_{sJ}(2632)$ was interpreted as the $J^P=1^-$ first radial excitation of $D^\ast_s(2112)$. The special decay manner was attributed to the node structure of the wave functions in Ref.~\cite{chao} or to the couple channels effect in Ref.~\cite{rupp}. However, the assignment of $D^+_{sJ}(2632)$ with $2^+$ $D_s(^3P_2)$ or $1^-$ $D_s(2^3S_1)$ was pointed out impossible in Refs.~\cite{zhang1,chang}, and the assignment of $D^+_{sJ}(2632)$ with $D_s(1^3D_1)$ was advocated in Refs.~\cite{chang,zhang2}. In the frame of heavy quark effective theory (HQET) sum rule~\cite{huang}, it was pointed out that $D^+_{sJ}(2632)$ was unlikely to be a conventional orbitally higher excited $c\bar{s}$ resonance.

$D^+_{sJ}(2632)$ was alternatively interpreted as a four-quark state in Refs.~\cite{Maiani,Y.-R,Yu-Qi}. The narrow width and exotic decay manner were interpreted in this assignment. However, the predicted partners and more decay channels of $D^+_{sJ}(2632)$ have not been observed. In Refs.~\cite{ted,swanson}, the authors  concluded that $D^+_{sJ}(2632)$ was an experimental artefact.

The $^3P_0$ model is a phenomenological method first put forth by Micu~\cite{micu1969} and subsequently developed by Yaouanc {\it et al.}~\cite{yaouanc1,yaouanc2,yaouanc3}. It has been employed successfully to explore Okubo-Zweig-Iizuka-allowed (OZI-allowed) hadronic decays of hadrons. Hadronic decays of the $J^P=0^-$ radial excitation $D_s$ with $M=2673$ MeV and $M=2650$ MeV have been explored in the $^3P_0$ model~\cite{gm2,gm}, but the hadronic production of this missing $D_s(2650)$ from higher excited resonances has not been explored. As that in Refs.~\cite{zhang3,zhang4}, it is interesting to explore the hadronic production of $D_s(2650)$ from higher excited resonances.

As for $D^+_{sJ}(2632)$, its hadronic decay has not yet been systematically studied in the $^3P_0$ model. In order to identify $D^+_{sJ}(2632)$, it is required to explore the hadronic decay features of $D^+_{sJ}(2632)$ no matter whether $D^+_{sJ}(2632)$ is an experimental artefact or not.

The paper is organized as follows. In the second section, the $^3P_0$ model and the parameters are briefly introduced. In Sec. III, the numerical results relevant to hadronic decays of $D_s(2650)$, $D_s(3P)$ and $D^+_{sJ}(2632)$ are presented. The summaries and discussions are presented in Sec. IV.

\section{$^3P_0$ MODEL}
Among models for hadronic decays, the $^3P_0$ model is popularly known as a quark-pair creation (QPC) model. Since the propose of $^3P_0$ model~\cite{micu1969,yaouanc1,yaouanc2,yaouanc3}, it has been extensively applied to mesons and baryons with success. To proceed with a practical computation, the formula in Refs.~\cite{zhang4,zhang5} are presented. In the $^3P_0$ model, the hadronic decay width of $A\to BC$ is
\begin{eqnarray}
\Gamma  = \pi ^2 \frac{|\vec{p}|}{M_A^2}\sum_{JL} |{\mathcal{M}^{JL}}|^2
\end{eqnarray}
where $\vec {p}$ is the momentum of the final mesons $B$ and $C$ in the initial meson $A$'s center-of-mass frame
\begin{eqnarray}
 |\vec{p}|=\frac{{\sqrt {[m_A^2-(m_B-m_C )^2][m_A^2-(m_B+m_C)^2]}}}{{2m_A }}
\end{eqnarray}
and $M^{JL}$ is the partial wave amplitude of $A \rightarrow BC$. The partial wave amplitude $M^{JL}$ is obtained from the helicity amplitude $\mathcal{M}^{M_{J_A } M_{J_B } M_{J_C }}$ in terms of the Jacob-Wick formula as follows
\begin{flalign}
\mathcal{M}^{JL} (A \to BC) &= \frac{{\sqrt {2L + 1} }}{{2J_A  + 1}} \nonumber \\
&\times\sum_{M_{J_B } ,M_{J_C } } \langle {L0JM_{J_A } } |{J_A M_{J_A } }\rangle  \nonumber \\
&\times\langle {J_B M_{J_B } J_C M_{J_C } } |J, {JM_{J_A } } \rangle \nonumber \\
 &\times \mathcal{M}^{M_{J_A } M_{J_B } M_{J_C } } (\vec{p})
\end{flalign}
where $\vec{J}=\vec{J_B}+\vec{J_C}$, $\vec{J_A}=\vec{J_B}+\vec{J_C}+\vec{L}$ and $M_{J_A}=M_{J_B}+M_{J_C}$. The helicity amplitude reads
\begin{flalign}
 &\delta ^3 (\vec{p}_B+ \vec{p}_C )\mathcal{M}^{M_{J_A } M_{J_B } M_{J_C }}\nonumber \\
 &=\sqrt {8E_A E_B E_C } \gamma \sum_{\mbox{\tiny$\begin{array}{c}
M_{L_A } ,M_{S_A } ,\\
M_{L_B } ,M_{S_B } ,\\
M_{L_C } ,M_{S_C } ,m\end{array}$}}  \langle {L_A M_{L_A } S_A M_{S_A } }| {J_A M_{J_A } }\rangle \nonumber \\
 &\times\langle L_B M_{L_B } S_B M_{S_B }|J_B M_{J_B } \rangle \langle L_C M_{L_C } S_C M_{S_C }|J_C M_{J_C }\rangle\nonumber \\
 & \times \langle {1m;1 - m}|{00} \rangle\langle \chi _{S_B M_{S_B }}^{13} \chi _{S_C M_{S_C } }^{24}|\chi _{S_A M_{S_A } }^{12} \chi _{1 - m}^{34}\rangle \nonumber \\
&\times\langle\varphi _B^{13} \varphi _C^{24}|\varphi _A^{12}\varphi _0^{34} \rangle I_{M_{L_B } ,M_{L_C } }^{M_{L_A },m} (\vec{p})
\end{flalign}
where $\vec{p}_B$ and $\vec{p}_C$ ($\vec{p}_B$=-$\vec{p}_C$=$\vec{p}$) are the momenta of final mesons $B$ and $C$ in $A$'s center-of-mass frame, respectively. $\gamma$ is a phenomenological parameter, which indicates the strength of the quark-pair creation from the vacuum. $I_{M_{L_B },M_{L_C}}^{M_{L_A },m}(\vec{p})$ is a momentum integral
\begin{flalign}
I_{M_{L_B } ,M_{L_C } }^{M_{L_A } ,m} (\vec{p})&= \int d \vec{k}_1 d \vec{k}_2 d \vec{k}_3 d \vec{k}_4 \nonumber \\
&\times\delta ^3 (\vec{k}_1 + \vec{k}_2-\vec{p}_A)\delta ^3 (\vec{k}_3+ \vec{k}_4)\nonumber \\
&\times \delta ^3 (\vec{p}_B- \vec{k}_1- \vec{k}_3 )\delta ^3 (\vec{p}_C- \vec{k}_2 -\vec{k}_4) \nonumber \\
& \times\Psi _{n_B L_B M_{L_B } }^* (\vec{k}_{13})\Psi _{n_cL_C  M_{L_c}}^* (\vec{k}_{24}) \nonumber \\
& \times \Psi _{n_A L_A M_{LA}} (\vec {k}_{12})Y _{1m}\left(\vec{k}_{34}\right)
\end{flalign}
with a relative momentum of quark $i$ and quark $j$: $\vec{k}_{ij}={{m_j\vec{k}_i-m_i\vec{k}_j}\over {m_i+m_j}}$ ($i,j=1,2,3,4$). $\Psi_{nLM_L}(\vec{k})$ are simple harmonic oscillator (SHO) wave functions for the mesons, and $y _{1m}(\vec{k})=|\vec{k}|Y_{1m}(\Omega)$ is the solid harmonic polynomial of the created P-wave quark pair.

In terms of $9j$ symbols~\cite{yaouanc3}, the flavor matrix element reads
\begin{flalign}
\langle\varphi _B^{13} \varphi _C^{24}|\varphi _A^{12}\varphi _0^{34} \rangle &= \sum_{I,I^3 } \langle {I_C I_C^3; I_B I^3_B }|{I_A, I_A^3 }\rangle \nonumber \\
&\times [(2I_B  + 1)(2I_C  + 1)(2I_A  + 1)]^{1/2} \nonumber \\
&\times \begin{Bmatrix}
    {I_1} & {I_3} & {I_B}\\
    {I_2} & {I_4} & {I_C}\\
    {I_A} &  {0}  & {I_A}\\ \end{Bmatrix}
\end{flalign}
where $I_i~(i=1,2,3,4)$ is the isospin of the four $u$, $d$, $s$ or $c$ quark. $I_A$, $I_B$ and $I_C$ are the isospins of the mesons $A$, $B$ and $C$, respectively.
$I^3_A$, $I^3_B$ and $I^3_C$ are the third isospin components of the mesons, and $\langle {I_C I_C^3; I_B I^3_B }|{I_A, I_A^3 }\rangle$ is the isospin matrix element.

The spin matrix element reads
\begin{flalign}
&\left\langle {\chi _{S_B M_{S_B } }^{13} \chi _{S_C M_{S_C } }^{24} } |{\chi _{S_A M_{S_A } }^{12} \chi _{1 - m}^{34} } \right\rangle \nonumber \\
&= (-1)^{S_C  + 1} [3(2S_B  + 1)(2S_C  + 1)(2S_A  + 1)]^{1/2} \nonumber \\
&\times\sum_{S,M_S } {\langle {S_B M_{S_B } S_C M_{S_C } }|{SM_S }\rangle \langle {SM_S }|{S_A M_{S_A } ;1, - m}\rangle}\nonumber \\
&\times
\begin{Bmatrix}
    {1/2} & {1/2} & {S_B}\\
    {1/2} & {1/2} & {S_C}\\
    {S_A} &  {1}  & {S}\\
\end{Bmatrix}.
\end{flalign}

\par With these formula in hand, we proceed with our calculation. In the practical calculation, the parameters have been employed as follows. The masses of the constituent quarks are taken to be, $m_c=1628$ MeV, $m_u=m_d=220$ MeV, and $m_s=419$ MeV~\cite{gm}. The masses and the effective scale parameters $\beta$ of relevant mesons are presented in Table~\ref{tab2}~\cite{pdg,gm}, where a common value $\beta=0.4$ GeV is employed for all light flavor mesons. For resonances included in PDG, their masses in PDG are employed (unlike those in Ref.~\cite{gm}), while for resonances not included in PDG or not observed by experiment, their masses and $\beta$ are taken from Ref.~\cite{gm}. $\gamma=6.95$ is employed in this paper, where $\gamma$ is $\sqrt{96\pi}$ times as the $\gamma=0.4$ adopted in Ref.~\cite{gm} for a different definition. The meson flavor follows the convention~\cite{gm}: $D^0=c\bar{u}$, $D^+=-c\bar{d}$, $D_s^+=-c\bar{s}$, $K^+=-u\bar{s}$, $K^-=s\bar{u}$, $\phi=-s\bar{s}$, and $\eta=(u\bar{u}-d\bar{d})/2-s\bar{s}/\sqrt{2}$.

\begin{table*}[t]\label{tab2}
\centering
\caption{Masses and $\beta$ values of mesons (MeV).}
\begin{tabular}{p{0.0cm}p{3.0cm}p{1.5cm}p{1.5cm}p{3.0cm}p{1.5cm}p{1.5cm}}
   \hline\hline
   & States &Mass &$\beta$ & States &Mass &$\beta$\\
   \hline
   &$(1^1S_0)K^{\pm\rm}$               & $493.677$     &$400 $     &$(1P_1)K_1(1270)^{0(\pm\rm)}$          &$1272$          &$400$    \\
   &$(1^1S_0)K^0$                      & $497.614$     &$ 400$     &$(1P_1^\prime)K_1(1400)^{0(\pm\rm)}$   &$1403$          &$400$           \\
   &$(1^3S_1)K^{*\pm\rm}$              & $891.66$      &$ 400$     &$(1^3P_2)K_2^*(1430)^0$                &$1432.4$        &$400$         \\
   &$(1^3S_1)K^{*0}$                   & $895.81$      &$400 $     &$(1^3P_2)K_2^*(1430)^{\pm\rm}$         &$1425.6$        &$400$             \\
   &$(1^1S_0)\eta$                     & $547.862$     &$ 400$     &$(1^3P_0)K_0^*(1430)^{0(\pm\rm)}$      &$1425$          &$400$      \\
   &$(1^3S_1)\phi$                     & $1019.461$    &$400 $     &$(1P_1)D_1(2430)^{0(\pm\rm)}$          &$2427$          &$475$ \\
   &$(1^1S_0)D^{\pm\rm}$               & $1869.61$     &$ 601$     &$(1P_1^\prime)D_1(2420)^{0}$           &$2421.4$        &$475$          \\
   &$(1^1S_0)D^0$                      & $1864.84$     &$601 $     &$(1P_1^\prime)D_1(2420)^{\pm\rm}$      &$2423.2$        &$475$         \\
   &$(1^3S_1)D^{*\pm\rm}$              & $2010.26$     &$ 516$     &$(1^3P_0)D_0(2400)^{0}$                &$2318$          &$516$   \\
   &$(1^3S_1)D^{*0}$                   & $2006.96$     &$ 516$     &$(1^3P_0)D_0(2400)^{\pm\rm}$           &$2403$          &$516$    \\
   &$(1^1S_0)D_s$                      & $1968.30$     &$651$      &$(1^3P_2)D_2(2460)^{\pm\rm}$           &$2464.3$        &$437$   \\
   &$(1^3S_1)D_S^*$                    & $2112.1$      &$562$      &$(1^3P_2)D_2(2460)^{0}$                &$2462.6$        &$437$    \\
   &$(2^1S_0)K(1460)^{0(\pm\rm)}$      & $1460$        &$400$      &$ (2^3S_1)D^*(2600)^0$                   &$2608.7$        &$434$   \\
   &$(2^3S_1)K^*(1410)^{0(\pm\rm)}$    & $1414$        &$400$      &$(2^3S_1)D^*(2600)^+$                    &$2621.3$        &$434$     \\
   &$(2^1S_0)D(2550)^{0(\pm\rm)}$      & $2539.4$      &$450$      &$(1^3D_1)D_1^*(2760)^0$                &$2763.3$        &$456$     \\
   &$(1^3P_0)D_{s0}(2317)$             & $2317.7$      &$542$      &$(1^3D_1)D_1^*(2760)^+$                &$2769.7$        &$456$     \\
   &$(1P_1)D_{s1}(2460)$               & $2459.5$      &$498$      &$(1^3D_3)D_3^*(2760)^0$                &$2763.3$        &$407$       \\
   &$(1P_1^\prime)D_{s1}(2536)$        & $2535.10$     &$498$      &$(1^3D_3)D_3^*(2760)^+$                &$2769.7$        &$407$   \\
   &$(1^3P_2)D_{s2}(2573)$             & $2571.9$      &$464 $     &$(1^1D_2)D(2750)^0 $                   &$2752.4$        &$428$  \\
   &$(2^3S_1)D^*_{s1}(2700)$           & $2708.3$      &$458 $     &$(1^1D_2)D(2750)^+ $                   &$2752.4$        &$428$    \\
   &$(1^3D_1)D^*_{s1}(2860)$           & $2859$        &$469 $     &$(1^3D_2)D(2750)^0$                    &$ 2752.4$       &$428$    \\
   &$(1^3D_3)D^*_{s3}(2860)$           & $2860.5$      &$426 $     &$(1^3D_2)D(2750)^+$                    &$2752.4$        &$428$  \\
   \hline\hline
\end{tabular}
\end{table*}

\section{Hadronic decays}

\subsection{$D_{s}(2650)$}

Two possible hadronic decay channels and partial decay widths of the $J^P=0^-$ radial excitation $D_s$ with $M=2650$ MeV ($D_s(2650)$) were calculated in the $^3P_0$ model~\cite{gm2}:
\begin{eqnarray}\nonumber
\Gamma(D_s(2^1S_0)\to D^*K)=78~\rm{MeV},\\\nonumber
\Gamma(D_s(2^1S_0)\to D^*_s\eta)=0.
\end{eqnarray}
Once the $J^P=0^-$ radial excitation $D_s$ has a mass $M=2673$ MeV, the partial decay widths become~\cite{gm2}:
\begin{eqnarray}\nonumber
\Gamma(D_s(2^1S_0)\to D^*K)=94~\rm{MeV},\\\nonumber
\Gamma(D_s(2^1S_0)\to D^*_s\eta)=0.6~\rm{MeV}.
\end{eqnarray}
In Ref.~\cite{gm}, the results  were improved as follows
\begin{eqnarray}\nonumber
\Gamma(D_s(2^1S_0)\to D^*K)=73.6~\rm{MeV},\\\nonumber
\Gamma(D_s(2^1S_0)\to D^*_s\eta)=0.
\end{eqnarray}
$D^*K$ decay channel is the dominant hadronic decay of $D_s(2650)$, which may be used as a typical evidence to identify the $J^P=0^-$ radial excitation $D_s$. $D(2550)$ was observed in inclusive $e^+e^-$ and $pp$ collisions, $D_s(2650)$ is naturally anticipated to be observed in these inclusive $e^+e^-$ and $pp$ collisions in $D^*K$ channel.

In other ways, $D_s(2650)$ can be produced from the hadronic decays of higher excited $D$ and $D_s$ mesons through $D_s(2650)K$ or $D_s(2650)\eta$ channels, respectively. Relevant hadronic partial decay widths have been estimated and presented in Table~\ref{tab3}.

\begin{table*}[t]
\centering
\caption{Hadronic partial decay widths of $D_s(2650)$ from highly excited resonances (MeV)}
\begin{tabular}{p{0cm}p{4.0cm}p{2.0cm}p{4.0cm}p{2.0cm}}
   \hline\hline
   & Mode& Width&  Mode & Width  \\
   \hline
   &$D(4^3S_1)\rightarrow D_s(2650)K$           &$0.1$                     &$ D_s(4^3S_1)\rightarrow D_s(2650)\eta$                &$0.0$            \\
   &$D(5^3S_1)\rightarrow D_s(2650)K$           &$0.0$                     &$ D_s(5^3S_1)\rightarrow D_s(2650)\eta$                &$0.0$            \\
   &$D(3^3P_0)\rightarrow D_s(2650)K$           &$1.8$                     &$D_s(3^3P_0)\rightarrow D_s(2650)\eta$                 &$0.1$           \\
   &$D(3^3P_2)\rightarrow D_s(2650)K$           &$2.8$                     &$D_s(3^3P_2)\rightarrow D_s(2650)\eta$                 &$2.4$         \\
   &$D(4^3P_0)\rightarrow D_s(2650)K$           &$0.2$                     &$D_s(4^3P_0) \rightarrow D_s(2650)\eta$                &$0.0$           \\
   &$D(4^3P_2) \rightarrow D_s(2650)K$          &$0.0$                     &$D_s(4^3P_2)\rightarrow D_s(2650)\eta$                 &$0.2$           \\
   &$D(2^3D_1)\rightarrow D_s(2650)K$           &$6.1$                     &$D_s(2^3D_1)\rightarrow D_s(2650)\eta$                 &$4.7$          \\
   &$D(2^3D_3)\rightarrow D_s(2650)K$           &$0.0$                     &$D_s(2^3D_3)\rightarrow D_s(2650)\eta$                 &$0.1$          \\
   &$D(3^3D_1)\rightarrow D_s(2650)K$           &$0.0$                     &$D_s(3^3D_1)\rightarrow D_s(2650)\eta$                 &$0.5$         \\
   &$D(3^3D_3)\rightarrow D_s(2650)K$           &$1.4$                     &$D_s(3^3D_3)\rightarrow D_s(2650)\eta$                 &$1.0$         \\
   &$D(1^3F_2)\rightarrow D_s(2650)K$           &$--$                      &$D_s(1^3F_2)\rightarrow D_s(2650)\eta$                 &$0.0$         \\
   &$D(2^3F_2)\rightarrow D_s(2650)K$           &$4.5$                     &$D_s(2^3F_2)\rightarrow D_s(2650)\eta$                 &$3.0$         \\
   &$D(2^3F_4)\rightarrow D_s(2650)K$           &$0.6$                     &$D_s(2^3F_4)\rightarrow D_s(2650)\eta$                 &$0.8$          \\
   &$D(1^3G_3)\rightarrow D_s(2650)K$           &$0.6$                     &$D_s(1^3G_3)\rightarrow D_s(2650)\eta$                 &$0.6$         \\
   &$D(1^3G_5)\rightarrow D_s(2650)K$           &$0.0$                     &$D_s(1^3G_5)\rightarrow D_s(2650)\eta$                 &$0.0$          \\
   &$D(2^3G_3)\rightarrow D_s(2650)K$           &$1.6$                     &$D_s(2^3G_3)\rightarrow D_s(2650)\eta$                 &$1.0$           \\
   &$D(2^3G_5)\rightarrow D_s(2650)K$           &$0.7$                     &$D_s(2^3G_5)\rightarrow D_s(2650)\eta$                 &$1.0$           \\
 \hline\hline
 \end{tabular}
 \label{tab3}
\end{table*}

\subsection{Hadronic decay of $D_s(3P)$ resonances}

$D_s(2650)$ can be produced from hadronic decays of higher excited $D$ and $D_s$ mesons. Hadronic decays of $D(2D)$ and $D_s(2D)$ have been explored in the $^3P_0$ model in Refs.~\cite{zhang3,zhang4}. Hadronic decays of $D_s(3P)$ have been studied in Ref.~\cite{gm}, where some decay channels were ignored. In this subsection, hadronic decays of $D_s(3P)$ resonances as an example are explored in detail.

In the $D_s(3P)$ multiplets, $D_s(3^3P_1)$ and $D_s(3^1P_1)$ may mix with each other. For lack of information, the detail of the mixing of $D_s(3^3P_1)$ with $D_s(3^1P_1)$ is not clear. Therefore, the OZI-allowed hadronic decay channels of $D_s(3^3P_0)$, $D_s(3^1P_1)$, $D_s(3^3P_1)$ and $D_s(3^3P_2)$ are studied separately.

The masses of $D_s(3P)$ are employed as follows~\cite{gm}: $M(D_s(3^3P_0))=3412$ MeV, $M(D_s(3^1P_1))=M(D_s(3^3P_1))=3425$ MeV, $M(D_s(3^3P_2))=3439$ MeV. The masses of $D_s(3^1P_1)$ and $D_s(3^3P_1)$ are assumed equal to the mean mass value of $D_s(3P_1)$ and $D_s(3P^\prime_1)$ in Ref.~\cite{gm}.
The final decay widths are given in Tables~\ref{tab6}-\ref{tab9}. All the four $D_s(3P)$ resonances have large total decay widths, which are larger than those in Ref.~\cite{gm}.

\begin{table*}[t]
\centering
\caption{Hadronic decay widths of $D_s(3^3P_0)$ (MeV).}
\centering
\begin{tabular}{p{0cm}p{4.0cm}p{2.0cm}p{4.0cm}p{2.0cm}}
   \hline\hline
   & Channels& Width&  Channels & Width  \\
   \hline
   &$D^0K^+$                    &$10.5$              &$D_1(2420)^+K^{0}$            &$0.1$           \\
   &$D^+K^0$                    &$10.3$              &$D_{s1}(2536)\eta$            &$0.3$          \\
   &$D_s\eta$                   &$0.2$               &$D_1(2430)^0K^{+}$            &$7.8$           \\
   &$D(2550)^0K^+$              &$9.8$               &$D_1(2430)^+K^{0} $           &$7.6$          \\
   &$D(2550)^+K^0$              &$9.7$               &$ D_{s1}(2460)\eta $          &$0.2$          \\
   &$D^0K(1460)^+$              &$5.5$               &$D_1(2420)^0K^{*+}$           &$8.1$          \\
   &$D^+K(1460)^0$              &$6.9$               &$D_1(2420)^+K^{*0}$           &$8.2$         \\
   &$D^{*0}K^{*+}$              &$3.2$               &$D_1(2430)^0K^{*+}$           &$5.7$          \\
   &$D^{*+}K^{*0}$              &$2.8$               &$D_1(2430)^+K^{*0}$           &$5.7$          \\
   &$D_s^*\phi$                 &$4.0$               &$ D^{*0}K_1(1270)^+$          &$1.1 $          \\
   &$D_0^{*}(2400)^0K^{*+} $    &$0.6$               &$D^{*+}K_1(1270)^0$           &$1.1$        \\
   &$ D_0^{*}(2400)^+K^{*0}$    &$2.2$               &$D^{*0}K_1(1400)^+$           &$0.3$        \\
   &$D_{s0}^{*}(2317)\phi$      &$2.0$               &$D_2^{*}(2460)^0K^{*+}$       &$3.0$          \\
   &$D^0K_1(1270)^+$            &$0.8$               &$ D_2^{*}(2460)^+K^{*0}$      &$ 2.5 $         \\
   &$D^+K_1(1270)^0$            &$0.6$               &$D(2750)^0K^{+} $             &$19.3$         \\
   &$D^0K_1(1400)^+$            &$1.5$               &$D(2750)^+K^{0}$              &$18.8$          \\
   &$D^+K_1(1400)^0$            &$1.7$               &$(2^1S_0)D_s(2650)\eta$       &$0.1$         \\
   &$D_1(2420)^0K^{+}$          &$0.1$               &$\Gamma_{total}$              &$162.3$          \\
   \hline\hline
   \end{tabular}
\label{tab6}
\end{table*}

\begin{table*}[t]
\centering
\caption{Hadronic decay widths of $D_s(3^1P_1)$ (MeV).}
\centering
\begin{tabular}{p{0cm}p{4.0cm}p{2.0cm}p{4.0cm}p{2.0cm}}
   \hline\hline
   & Channels& Width&  Channels & Width  \\
   \hline
   &$D^{0}K^{*+}$                      &$3.9$                     &$ D_{s1}(2460)\eta$                 &$0.0$            \\
   &$D^{+}K^{*0}$                      &$3.7$                     &$ D_1(2420)^0K^{*+}$                &$7.9$            \\
   &$D_s\phi$                          &$0.1$                     &$D_1(2420)^+K^{*0}$                 &$8.1$           \\
   &$D^{0}K^*(1410)^+$                 &$13.0$                    &$D_1(2430)^0K^{*+}$                 &$9.3$         \\
   &$D^{+}K^*(1410)^0$                 &$12.4$                    &$ D_1(2430)^+K^{*0}$                &$9.3$           \\
   &$ D^{*0}K^+$                       &$7.7$                     &$D^{*0}K_1(1270)^+$                 &$8.1$           \\
   &$D^{*+}K^0 $                       &$7.6$                     &$D^{*+}K_1(1270)^0$                 &$8.4$          \\
   &$D_s^*\eta$                        &$0.2$                     &$ D^{*0}K_1(1400)^+$                &$2.6$          \\
   &$D^{*0}K^{*+}$                     &$2.7$                     &$D^{*+}K_1(1400)^0$                 &$1.8$         \\
   &$D^{*+}K^{*0}$                     &$2.4$                     &$D^{0}K_2^{*}(1430)^+$              &$3.1$         \\
   &$D_s^*\phi $                       &$2.0$                     &$D^{+}K_2^{*}(1430)^0$              &$3.2$         \\
   &$D^{*0}K^*(1410)^+ $               &$15.9$                    &$D_2^{*}(2460)^0K^{+}$              &$14.6$         \\
   &$D^{*+}K^*(1410)^0$                &$4.5$                     &$D_2^{*}(2460)^+K^{0}$              &$14.4$          \\
   &$D_0^{*}(2400)^0K^{+}$             &$2.5$                     &$D_{s2}^*(2573)\eta$                &$3.9$         \\
   &$D_0^{*}(2400)^+K^{0}$             &$1.3$                     &$D_2^{*}(2460)^0K^{*+}$             &$13.0$          \\
   &$D_{s0}^{*}(2317)\eta$             &$0.2$                     &$D_2^{*}(2460)^+K^{*0}$             &$12.4$           \\
   &$D^{0}K_0^*(1430)^+$               &$1.4$                     &$D^*(2600)^0K^{+}$                  &$11.9$           \\
   &$D^{+}K_0^*(1430)^0$               &$1.6$                     &$D^*(2600)^+K^{0}$                  &$12.0$           \\
   &$D_0^{*}(2400)^0K^{*+}$            &$0.0$                     &$D(2750)^0K^{+}$                    &$0.0$           \\
   &$D_0^{*}(2400)^+K^{*0}$            &$0.0$                     &$D(2750)^+K^{0}$                    &$0.0$          \\
   &$D_{s0}^{*}(2317)\phi$             &$0.0$                     &$D_1^*(2760)^0K^{+}$                &$0.7$          \\
   &$D^0K_1(1270)^+$                   &$0.0$                     &$D_1^*(2760)^+K^{0} $               &$0.7$           \\
   &$D^+K_1(1270)^0$                   &$0.0$                     &$D_3^*(2760)^0K^{+}$                &$14.4$           \\
   &$D^0K_1(1400)^+$                   &$0.0$                     &$D_3^*(2760)^+K^{0}$                &$13.1$          \\
   &$D^+K_1(1400)^0$                   &$0.0$                     &$D_{s1}^*(2860)\eta$                &$0.5$          \\
   &$D_1(2420)^0K^{+}$                 &$0.0$                     &$D_{s3}^*(2860)\eta $               &$0.0$        \\
   &$D_1(2420)^+K^{0}$                 &$0.0$                     &$D_{s1}^*(2700)\eta$                &$3.5$           \\
   &$D_{s1}(2536)\eta$                 &$0.0$                     &$\Gamma_{total}$                    &$260.4$           \\
   &$D_1(2430)^0K^{+}$                 &$0.0$                     &                                    &          \\
   &$D_1(2430)^+K^{0}$                 &$0.4$                     &                                    &          \\

   \hline\hline
   \end{tabular}
\label{tab7}
\end{table*}

\begin{table*}[t]
\centering
\caption{Hadronic decay widths of $D_s(3^3P_1)$ (MeV).}
\centering
\begin{tabular}{p{0cm}p{4.0cm}p{2.0cm}p{4.0cm}p{2.0cm}}
   \hline\hline
   & Channels& Width&  Channels & Width  \\
   \hline
   &$D^{0}K^{*+}$                      &$4.1$                     &$ D_{s1}(2460)\eta$                 &$0.0$            \\
   &$D^{+}K^{*0}$                      &$3.8$                     &$ D_1(2420)^0K^{*+}$                &$4.5$            \\
   &$D_s\phi$                          &$0.2$                     &$D_1(2420)^+K^{*0}$                 &$4.6$           \\
   &$D^{0}K^*(1410)^+$                 &$7.2$                     &$D_1(2430)^0K^{*+}$                 &$11.9$         \\
   &$D^{+}K^*(1410)^0$                 &$6.6$                     &$ D_1(2430)^+K^{*0}$                &$12.0$           \\
   &$ D^{*0}K^+$                       &$8.5$                     &$D^{*0}K_1(1270)^+$                 &$7.6$           \\
   &$D^{*+}K^0 $                       &$8.3$                     &$D^{*+}K_1(1270)^0$                 &$8.0$          \\
   &$D_s^*\eta$                        &$0.1$                     &$ D^{*0}K_1(1400)^+$                &$2.3$          \\
   &$D^{*0}K^{*+}$                     &$3.7$                     &$D^{*+}K_1(1400)^0$                 &$1.7$         \\
   &$D^{*+}K^{*0}$                     &$3.3$                     &$D^{0}K_2^{*}(1430)^+$              &$2.5$         \\
   &$D_s^*\phi $                       &$2.0$                     &$D^{+}K_2^{*}(1430)^0$              &$2.8$         \\
   &$D^{*0}K^*(1410)^+ $               &$0.0$                     &$D_2^{*}(2460)^0K^{+}$              &$9.1$         \\
   &$D^{*+}K^*(1410)^0$                &$0.0$                     &$D_2^{*}(2460)^+K^{0}$              &$9.0$          \\
   &$D_0^{*}(2400)^0K^{+}$             &$0.0$                     &$D_{s2}^*(2573)\eta$                &$2.6$         \\
   &$D_0^{*}(2400)^+K^{0}$             &$0.0$                     &$D_2^{*}(2460)^0K^{*+}$             &$3.6$          \\
   &$D_{s0}^{*}(2317)\eta$             &$0.0$                     &$D_2^{*}(2460)^+K^{*0}$             &$3.2$           \\
   &$D^{0}K_0^*(1430)^+$               &$0.0$                     &$D^*(2600)^0K^{+}$                  &$9.8$           \\
   &$D^{+}K_0^*(1430)^0$               &$0.0$                     &$D^*(2600)^+K^{0}$                  &$9.3$           \\
   &$D_0^{*}(2400)^0K^{*+}$            &$0.3$                     &$D(2750)^0K^{+}$                    &$6.7$           \\
   &$D_0^{*}(2400)^+K^{*0}$            &$1.7$                     &$D(2750)^+K^{0}$                    &$6.5$          \\
   &$D_{s0}^{*}(2317)\phi$             &$1.8$                     &$D_1^*(2760)^0K^{+}$                &$0.1$          \\
   &$D^0K_1(1270)^+$                   &$0.0$                     &$D_1^*(2760)^+K^{0} $               &$0.1$           \\
   &$D^+K_1(1270)^0$                   &$0.0$                     &$D_3^*(2760)^0K^{+}$                &$12.0$           \\
   &$D^0K_1(1400)^+$                   &$1.2$                     &$D_3^*(2760)^+K^{0}$                &$11.0$          \\
   &$D^+K_1(1400)^0$                   &$1.4$                     &$D_{s1}^*(2860)\eta$                &$0.3$          \\
   &$D_1(2420)^0K^{+}$                 &$1.7$                     &$D_{s3}^*(2860)\eta $               &$0.0$        \\
   &$D_1(2420)^+K^{0}$                 &$1.6$                     &$D_{s1}^*(2700)\eta$                &$2.3$           \\
   &$D_{s1}(2536)\eta$                 &$0.2$                     &$\Gamma_{total}$                    &$204.8$           \\
   &$D_1(2430)^0K^{+}$                 &$1.8$                     &                                    &           \\
   &$D_1(2430)^+K^{0}$                 &$1.8$                     &                                    &         \\

   \hline\hline
   \end{tabular}
\label{tab8}
\end{table*}

\begin{table*}[t]
\centering
\caption{Hadronic decay widths of $D_s(3^3P_2)$ (MeV).}
\centering
\begin{tabular}{p{0cm}p{4.0cm}p{2.0cm}p{4.0cm}p{2.0cm}}
   \hline\hline
   & Channels& Width&  Channels & Width  \\
   \hline
   &$D^0K^+$                        &$1.0$                      &$D^{*0}K_0^*(1430)^+$               &$0.3$              \\
   &$D^+K^0$                        &$1.1$                      &$D^{*+}K_0^*(1430)^0$               &$0.2$             \\
   &$D_s\eta$                       &$0.3$                      &$D_1(2420)^0K^{*+}$                 &$4.1$             \\
   &$D(2550)^0K^+$                  &$2.6$                      &$D_1(2420)^+K^{*0}$                 &$4.3$              \\
   &$D(2550)^+K^0$                  &$2.7$                      &$ D_1(2430)^0K^{*+}$                &$3.1$             \\
   &$D^0K(1460)^+$                  &$5.3$                      &$D_1(2430)^+K^{*0}$                 &$3.0$            \\
   &$D^+K(1460)^0$                  &$5.0$                      &$D^{*0}K_1(1270)^+$                 &$3.1$            \\
   &$D^{0}K^{*+}$                   &$2.3$                      &$ D^{*+}K_1(1270)^0$                &$3.2$            \\
   &$D^{+}K^{*0}$                   &$2.2$                      &$D^{*0}K_1(1400)^+$                 &$1.8$            \\
   &$D_s\phi$                       &$0.0$                      &$D^{*+}K_1(1400)^0$                 &$1.6$            \\
   &$D^{0}K^*(1410)^+$              &$11.1$                     &$D^{0}K_2^{*}(1430)^+$              &$1.1$           \\
   &$D^{+}K^*(1410)^0$              &$10.9$                     &$D^{+}K_2^{*}(1430)^0$              &$1.0$           \\
   &$D(2550)^0K^{*+}$               &$0.0$                      &$D_2^{*}(2460)^0K^{+}$              &$6.5$            \\
   &$D(2550)^+K^{*0}$               &$0.0$                      &$D_2^{*}(2460)^+K^{0}$              &$6.5$             \\
   &$D^{*0}K^+$                     &$3.8$                      &$D_{s2}^*(2573)\eta$                &$1.8$            \\
   &$D^{*+}K^0$                     &$3.8$                      &$D^{*0}K_2^*(1430)^+$               &$1.6$            \\
   &$D_s^*\eta$                     &$0.3$                      &$D_2^{*}(2460)^0K^{*+}$             &$25.0$             \\
   &$ D^{*0}K^{*+}$                 &$4.8$                      &$D_2^{*}(2460)^+K^{*0}$             &$24.9$            \\
   &$ D^{*+}K^{*0}$                 &$4.4$                      &$D^*(2600)^0K^{+}$                  &$7.1$             \\
   &$D_s^*\phi$                     &$1.7$                      &$D^*(2600)^+K^{0}$                  &$7.6$            \\
   &$D^{*0}K^*(1410)^+$             &$45.2$                     &$D(2750)^0K^{+}$                    &$2.0$            \\
   &$D^{*+}K^*(1410)^0$             &$44.5$                     &$D(2750)^+K^{0} $                   &$2.0$            \\
   &$D_0^{*}(2400)^0K^{*+}$         &$0.1$                      &$D(2750)^0K^{+} $                   &$0.4$            \\
   &$D_0^{*}(2400)^+K^{*0}$         &$1.1$                      &$D(2750)^+K^{0}$                    &$0.4$            \\
   &$D_{s0}^{*}(2317)\phi$          &$1.3$                      &$D_1^*(2760)^0K^{+}$                &$0.2$            \\
   &$D^0K_1(1270)^+$                &$2.9$                      &$D_1^*(2760)^+K^{0}$                &$0.2$           \\
   &$D^+K_1(1270)^0$                &$2.7$                      &$D_3^*(2760)^0K^{+}$                &$4.7$            \\
   &$D^0K_1(1400)^+$                &$1.6$                      &$D_3^*(2760)^+K^{0}$                &$4.3$             \\
   &$D^+K_1(1400)^0$                &$1.5$                      &$D_s(2650)\eta$                     &$2.4$              \\
   &$D_1(2420)^0K^{+}$              &$1.9$                      &$D_{s1}^*(2860)\eta$                &$0.0$              \\
   &$D_1(2420)^+K^{0}$              &$1.8$                      &$D_{s3}^*(2860)\eta$                &$0.0$              \\
   &$D_{s1}(2536)\eta$              &$0.1$                      &$D_{s1}^*(2700)\eta$                &$3.0$              \\
   &$D_1(2430)^0K^{+}$              &$4.9$                      &$\Gamma_{total}$                    &$307.9$             \\
   &$D_1(2430)^+K^{0}$              &$4.9$                      &                                    &              \\
   &$D_{s1}(2460)\eta$              &$2.7$                      &                                    &              \\
  \hline\hline
   \end{tabular}
\label{tab9}
\end{table*}

From Table~\ref{tab6}, $\Gamma_{total}(D_s(3^3P_0))=162.3$ MeV and the dominant decay channels of $D_s(3^3P_0)$ are: $(1^1D_2)D(2750)K$, $DK$, $D(2550)K$ and $D_1(2420)K^*$.

From Table~\ref{tab7}, $\Gamma_{total}(D_s(3^1P_1))=260.4$ MeV and the dominant decay channels of $D_s(3^1P_1)$ are: $D^*_2(2460)K$, $D_3^*(2760)K$ , $DK^*(1410)$ and $D^*_2(2460)K^*$.

From Table~\ref{tab8}, $\Gamma_{total}(D_s(3^3P_1))=204.8$ MeV and the dominant decay channels of $D_s(3^3P_1)$ are: $D_1(2430)K^*$, $D_3^*(2760)K$, $D(2600)K$ and $D^*_2(2460)K$.

From Table~\ref{tab9}, $\Gamma_{total}(D_s(3^3P_2))=307.9$ MeV and the dominant decay channels of $D_s(3^3P_2)$ are: $D^*K^*(1410)$, $D^*_2(2460)K^*$, $DK^*(1410)$ and $D(2600)K$.

\subsection{$D^+_{sJ}(2632)$}

In Table I, $D^*_{s1}(2700)$ and $D^*_{s3}(2860)$ are tentatively identified with $D_s(2^3S_1)$ and $D_s(1^3D_1)$, respectively. However, as introduced in the first section, $D^+_{sJ}(2632)$ was once regarded as a $D_s(2^3S_1)$, $D_s(1^3D_1)$ or their admixture if its lower mass is ignored. In order to have a look at these possibilities, let's explore the hadronic decay of $D^+_{sJ}(2632)$. The mixing of $D_s(2^3S_1)$ and $D_s(1^3D_1)$ will not be taken into account for its unclear mixing feature.

In $D_s(2^3S_1)$ or $D_s(1^3D_1)$ assignment, $D^+_{sJ}(2632)$ may decay into $D^0K^+$, $D^+K^0$, $D_s\eta$, $D^{*0}K^+$ and $D^{*+}K^0$ final states. The partial and total decay widths are given in Table~\ref{tab4}.

\begin{table*}[t]
\centering
\caption{Partial and total widths of $D^+_{sJ}(2632)$ (MeV) and $\Gamma(D^0K^+)/\Gamma(D^+_s\eta)$ in different assignments.}

\begin{tabular}{p{0cm}p{3.2cm}p{1.2cm}p{1.2cm}p{1.2cm}p{1.2cm}p{1.2cm}p{1.2cm}p{1.2cm}}
   \hline\hline
    &$Assignment$                      &$D^0K^+$   &$D^+K^0$    &$D_s\eta$    &$D^{*0}K^+$    &$D^{*+}K^0$    &$\Gamma_{total}$   &$\frac{\Gamma(D^0K^+)}{\Gamma(D^+_s\eta)}$\\
    \hline
    &$D^+_{sJ}(2632)(2^3S_1)$                     &$17.8$     &$17.5$     &$3.5$      &$22.9$        &$21.6$        &$83.2$           &$5.02$\\
    &$D^+_{sJ}(2632)(1^3D_1)$                     &$34.2$     &$33.4$     &$4.3$      &$8.8$         &$8.3$         &$88.9$           &$7.90$\\
\hline\hline
\end{tabular}
\label{tab4}
\end{table*}

In the case of a $D_s(2^3S_1)$, we have the following ratios:
\begin{eqnarray}
\frac{\Gamma(DK)}{\Gamma(D^*K)}=0.79;~~~~\frac{\Gamma(D^+_s\eta)}{\Gamma(D^*K)}=0.08.
\end{eqnarray}

In the case of a $D_s(1^3D_1)$, we have the following ratios:
\begin{eqnarray}
\frac{\Gamma(DK)}{\Gamma(D^*K)}=3.95;~~~~\frac{\Gamma(D^+_s\eta)}{\Gamma(D^*K)}=0.25.
\end{eqnarray}
Obviously, the decay manner of $D^+_{sJ}(2632)$ is different in different assignments. In fact, this decay manner can be employed to distinguish $D_s(2^3S_1)$ from $D_s(1^3D_1)$.

In particular, the estimated ratios ${\Gamma(D^0K^+)\over\Gamma(D^+_s\eta)}$, $5.02$ and $7.90$, are much larger than the observed $0.14\pm 0.06$. Obviously, $D^+_{sJ}(2632)$ cannot be a conventional charmed strange $c\bar s$ meson.

\section{CONCLUSIONS AND DISCUSSIONS}

\par The missing $J^P=0^-$ radial excitation $D_s(2650)$ is anticipated to have mass $2650$ MeV. $D^*K$ channel is the dominant hadronic decay channel of $D_s(2650)$, and $D_s(2650)$ is anticipated to be observed in inclusive $e^+e^-$ and $pp$ collisions in $D^*K$ channel. $D_s(2650)$ can be produced from higher excited $D$ through $D_s(2650)K$ channel or from $D_s$ meson through $D_s(2650)\eta$ channel.

Hadronic decays of $D_s(3P)$ have been studied. All the hadronic decay channels are given, and the decay widths have been estimated. The four $D_s(3P)$ resonances have large total decay widths. The dominant decay channels of $D_s(3^3P_0)$ are: $(1^1D_2)D(2750)K$, $DK$, $D(2550)K$ and $D_1(2420)K^*$. The dominant decay channels of $D_s(3^1P_1)$ are: $D^*_2(2460)K$, $D_3^*(2760)K$ , $DK^*(1410)$ and $D^*_2(2460)K^*$. The dominant decay channels of $D_s(3^3P_1)$ are: $D_1(2430)K^*$, $D_3^*(2760)K$, $D(2600)K$ and $D^*_2(2460)K$. The dominant decay channels of $D_s(3^3P_2)$ are: $D^*K^*(1410)$, $D^*_2(2460)K^*$, $DK^*(1410)$ and $D(2600)K$.

$D^+_{sJ}(2632)$ has a mass comparable to those of predicted $D_s(2^1S_0)$, $D_s(2^3S_1)$ and $D_s(1^3D_1)$. However, $D^+_{sJ}(2632)$ has an exotic ratio ${\Gamma(D^0K^+)\over\Gamma(D^+_s\eta)}=0.14\pm 0.06$, which is much smaller than the predicted ones in these possibilities. In short, $D^+_{sJ}(2632)$ cannot be a conventional charmed strange $c\bar s$ meson, which keeps rooms for $D_s(2650)$, $D^*_{s1}(2700)$ and $D^*_{s1}(2860)$.

\begin{acknowledgments}
This work is supported by National Natural Science Foundation of China under the grants: 11475111 and 11075102. It is also supported by the Innovation Program of Shanghai Municipal Education Commission under the grant No. 13ZZ066.
\end{acknowledgments}


\begin{thebibliography}{100}
\bibitem{pdg}
C. Patrignani {\it et al}. (Particle Data Group), Chin. Phys. C, {\bf 40}, 100001 (2016).
\bibitem{gi}
S. Godfrey and N. Isgur, Phys. Rev. D {\bf 32}, 189 (1985).
\bibitem{amo}
P.del Amo Sanchez {\it et al.} (BABAR Collaboration), Phys. Rev. D 82, 111101 (2010).
\bibitem{lhcb1}
R. Aaij {\it et al.} (LHCb collaboration), J. High Energy Phys. 09, 145 (2013).
\bibitem{lhcb2}
R. Aaij {\it et al.} (LHCb collaboration), Phys. Rev. D {\bf 91}, 092002 (2015).
\bibitem{lhcb3}
R. Aaij {\it et al.} (LHCb collaboration), Phys. Rev. D {\bf 92}, 032002 (2015).
\bibitem{belle}
J. Brodzicka, {\it et al.} (Belle Collaboration), Phys. Rev. Lett. 100, 092001 (2008).
\bibitem{babar}
B. Aubert, {\it et al.} (BaBar Collaboration), Phys. Rev. D {\bf 80}, 092003 (2009).
\bibitem{lhcb4}
R. Aaij {\it et al.} (LHCb collaboration), Phys. Rev. D {\bf 90}, 072003 (2014).
\bibitem{lhcb5}
R. Aaij et al. (LHCb collaboration), Phys. Rev. Lett. {\bf 113}, 162001 (2014).
\bibitem{colangelo}
P. Colangelo, F. De Fazio, F. Giannuzzi and S. Nicotri, Phys. Rev. D {\bf 86}, 054024 (2012).
\bibitem{gm}
S. Godfrey and K. Moats, Phys Rev. D {\bf 93}, 034035 (2016).
\bibitem{liu3}
Zhi-Feng Sun, Jie-Sheng Yu, Xiang Liu and Takayuki Matsuki, Phys. Rev. D {\bf 82}, 111501 (R) (2010).
\bibitem{zhong}
Xian-Hui Zhong, Phys. Rev. D {\bf 82}, 114014 (2010).
\bibitem{wang}
Zhi-Gang Wang, Phys. Rev. D {\bf 83}, 014009 (2011).
\bibitem{chen2}
Bing Chen, Ling Yuan and Ailin Zhang, Phys. Rev. D {\bf 83}, 114025 (2011).
\bibitem{chen}
Bing Chen, Xiang Liu and Ailin Zhang, Phys. Rev. D {\bf 92}, 034005 (2015).
\bibitem{zhang3}
Jing Ge, Dan-Dan Ye and Ailin Zhang, Eur. Phys. J. C {\bf75}, 178 (2015).
\bibitem{zhang4}
Ze Zhao, Yu Tian and Ailin Zhang, Phys. Rev. D {\bf 94}, 114035 (2016).
\bibitem{zhu}
Hua-Xing Chen, Wei Chen, Xiang Liu, Yan-Rui Liu and Shi-Lin Zhu, arXiv:1609.08928.
\bibitem{SELEX}
A.V. Evdokimov {\it et al.} (SELEX Collaboration), Phys. Rev. Lett, {\bf 93}, 242001 (2004).
\bibitem{babar2}
B. Aubert {\it et al.} (BABAR Collaboration), arXiv: hep-ex/0408087.
\bibitem{chao}
Kuang-Ta Chao, Phys. Lett. B {\bf 599}, 43 (2004).
\bibitem{ted}
T. Barnes, F.E. Close, J.J. Dudek, S. Godfrey, and E.S. Swanson, Phys. Lett. B {\bf 600}, 223 (2004).
\bibitem{rupp}
E.van Beveren and George Rupp, Phys. Rev. Lett. {\bf 93}, 202001 (2004).
\bibitem{zhang1}
Ailin Zhang, Phys. Rev. D {\bf 72}, 017902 (2005).
\bibitem{chang}
Chao-Hsi Chang, C. S. Kim and Guo-Li Wang, Phys. Lett. B {\bf 623}, 218 (2005).
\bibitem{zhang2}
Bing Chen, Deng-Xia Wang and Ailin Zhang, Phys. Rev. D {\bf 80}, 071502 (2009).
\bibitem{huang}
Ming-Qiu Huang and Dao-Wei Wang, Phys. Rev. D {\bf 71}, 114015 (2005).
\bibitem{Y.-R}
Y.-R. Liu, Shi-Lin Zhu, Y.-B. Dai and C. Liu, Phys. Rev. D {\bf 70}, 094009 (2004).
\bibitem{Maiani}
L. Maiani, F. Piccinini, A. D. Polosa and V. Riquer, Phys. Rev. D {\bf 70}, 054009 (2004).
\bibitem{Yu-Qi}
Yu-Qi Chen and Xue-Qian Li, Phys. Rev. Lett {\bf 93}, 232001 (2004).
\bibitem{swanson}
E.S. Swanson, Phys. Rept. {\bf 429}, 243 (2006).
\bibitem{micu1969}
L. Micu, Nucl. Phys. B {\bf 10}, 521 (1969).
\bibitem{yaouanc1}
A. Le Yaouanc, L. Oliver, O. P$\grave{e}$ne and J.C. Raynal, Phys. Rev. D {\bf 8}, 2223 (1973); {\bf 9}, 1415 (1974); {\bf 11}, 1272 (1975).
\bibitem{yaouanc2}
A. Le Yaouanc, L. Oliver, O. P$\grave{e}$ne and J.C. Raynal, Phys. Lett. B {\bf 71}, 397 (1977); {\bf 72}, 57 (1977).
\bibitem{yaouanc3}
A. Le Yaouanc, L. Oliver, O. P$\grave{e}$ne and J.C. Raynal, {\it Hadron Transitions in the Quark Model} (Gordon and Breach Science Publishers, New York, 1987).
\bibitem{gm2}
S. Godfrey and K. Moats, Phys Rev. D {\bf 90}, 117501 (2014).
\bibitem{zhang5}
Ze Zhao, Dan-Dan Ye and Ailin Zhang, Phys. Rev. D {\bf 94}, 114020 (2016).

\end{thebibliography}
\end{document}